\documentclass[twocolumn,superscriptaddress]{revtex4-2}
\usepackage{amsmath,amssymb,amsfonts}
\usepackage{graphicx}
\usepackage{hyperref}
\usepackage{bbold}
\usepackage{bm}
\usepackage{cancel}
\usepackage{verbatim}
\usepackage[caption=false]{subfig}
\usepackage{float}

\newcommand{\psit}{\psi_{\bm\theta}}
\newcommand{\psidt}{\psi_{\bm\theta+d\bm\theta}}
\newcommand{\deriv}[2]{\frac{\partial #1}{\partial #2}}
\newcommand{\LRangle}[1]{\left\langle#1\right\rangle}

\begin{document}

\title{Natural Gradient Optimization for Optical Quantum Circuits}

\author{Yuan Yao}
\email{yuan.yao@telecom-paris.fr}
\affiliation{Institut Polytechnique de Paris}
\affiliation{T\'el\'ecom Paris, LTCI, 20 Place Marguerite Peray, 91120 Palaiseau, France}

\author{Pierre Cussenot}
\affiliation{Institut Polytechnique de Paris}
\affiliation{Physics Department, \'Ecole Polytechnique, 91120 Palaiseau, France}

\author{Richard A. Wolf}
\affiliation{Institut Polytechnique de Paris}
\affiliation{T\'el\'ecom Paris, LTCI, 20 Place Marguerite Peray, 91120 Palaiseau, France}

\author{Filippo Miatto}
\affiliation{Xanadu, Toronto, ON, M5G 2C8, Canada}
\affiliation{Institut Polytechnique de Paris}
\affiliation{T\'el\'ecom Paris, LTCI, 20 Place Marguerite Peray, 91120 Palaiseau, France}

\begin{abstract}
Optical quantum circuits can be optimized using gradient descent methods, as the gates in a circuit can be parametrized by continuous parameters. However, the parameter space as seen by the cost function is not Euclidean, which means that the Euclidean gradient does not generally point in the direction of steepest ascent. In order to retrieve the steepest ascent direction, it is possible to take into account the local metric in the space parameters in what is known as the Natural Gradient (NG) method. In this work we implement the Natural Gradient (NG) for optical quantum circuits. In particular, we adapt the NG approach to a complex-valued parameter space. We then compare the NG approach to vanilla gradient descent and to Adam over two state preparation tasks: a single-photon source and a Gottesman-Kitaev-Preskill state source. We observe that the NG approach converges faster (due in part to the possibility of using larger learning rates) and with a significantly smoother decay of the cost function throughout the optimization.
\end{abstract}

\maketitle

\section{Introduction}

Optical quantum circuits are at the basis of continuous-variable (CV) quantum computation  \cite{Killoran_2019_continuous, Arrazola_2019}. Instead of qubits, which are defined in a two-dimensional Hilbert space, CV circuits work with optical modes, which are defined in an infinite-dimensional Hilbert space. For instance, a particular class of CV circuits has a layered architecture inspired by classical neural networks, where the layers alternate between Gaussian (akin to an affine linear transformation) and non-Gaussian (akin to an activation function) \cite{Killoran_2019_continuous}. Gaussian layers can be implemented by combining building blocks such as beam-splitters, phase-shifters (which are parametrized by real parameters), single-mode squeezers and single-mode displacements (which are parametrized by complex parameters). The non-Gaussian layers can be implemented by any non-linear gate such as the Kerr interaction (which is parametrized by a real parameter), although purely optical cubic or higher-order nonlinear optical interactions are being developed \cite{boyd2020nonlinear}. We should point out that non-linear gates are extremely challenging to implement in practice, and that alternatively one could use Gaussian operations with non-Gaussian resource states or non-Gaussian measurements as a source of non-linearity \cite{bourassa2021fast}. 

CV circuits may be used to prepare target states (state synthesis) or to approximate given quantum gates (gate synthesis). One can then set up a variational procedure to find the best set of parameters for the circuit to perform as well as possible with respect to a given cost function \cite{Arrazola_2019}. Several optimization algorithms can be used to perform cost function minimization, among which we find gradient-based optimizers.

The Natural Gradient (NG) has long been established as an approach to learning tasks that outperforms vanilla gradient descent, at the expense of needing a matrix inversion of an $n\times n$ matrix at each step, where $n$ is the number of parameters being updated during that step \cite{amari1996neural, amari1998natural, amari1998why}. Such matrix represents the metric tensor of the parameter space as seen by the cost function, and its purpose is to find the steepest descent direction while taking into account the local geometry \cite{amari1998natural, amari1998why, yamamoto2019}. The NG approach has also been proposed as part of variational quantum algorithms \cite{haug2021optimal}.

In the context of parametrized quantum systems, the metric tensor for pure states is the Fubini-Study metric tensor \cite{Stokes2020}, which can then be adapted for density matrices \cite{koczor2020quantum}. The quantum NG has proven successful in boosting hybrid quantum-classical algorithms such as Variational Quantum Eigensolvers (VQEs) for qubit-systems \cite{Stokes2020, koczor2020quantum, wierichs2020avoiding} outperforming other optimization methods \cite{yamamoto2019, wierichs2020avoiding}. In fact, the quantum NG yields stable results for various system sizes and it achieves convergence in fewer epochs than Euclidean gradient approaches \cite{Stokes2020, koczor2020quantum, wierichs2020avoiding}. In the VQE scenario, the metric tensor may be evaluated with the gradient directly on quantum hardware \cite{koczor2020quantum, Schuld_2019, Yuan_2019, Li_2017_efficient} and then inverted classically. 

The quantum NG looks also promising in the Noisy Intermediate-Scale Quantum (NISQ \cite{Preskill_2018}) computing paradigm, as it helps with noisy measurement situations \cite{koczor2020quantum}. Furthermore, some block-diagonal approximations have been introduced for layered quantum circuits \cite{Stokes2020}, which can reduce the computational load without excessively compromising the advantages. Note however that the metric tensor used for quantum states can be non-invertible or ill-conditioned \cite{yamamoto2019, Straaten2020measurement, wierichs2020avoiding}. This corresponds to cases where one or more parameters become redundant, which means that a single quantum state corresponds to multiple parameter values. To avoid this, one can introduce regularizations \cite{Straaten2020measurement, Stokes2020} and use the Penrose pseudo-inverse rather than the inverse of the metric \cite{Bromley_2020}.

In the case of qubit-based circuits, the gates are parametrized by real parameters. In contrast, CV circuits are more naturally parametrized by a combination of real and complex parameters. In our work we propose an implementation of QNG for optical quantum circuits that can involve complex parameters. We present numerical simulations to compare our implementation of QNG to SGD and Adam, while making use of a recent algorithm to recursively compute gradients in optical circuits \cite{miatto2020fast, yuan2021fast} in order to bypass some dimension cutoff issues \cite{Arrazola_2019}.

This article is organized as follows:
\begin{itemize}
    \item In section \ref{Previously}, we recall how the Natural Gradient is introduced on Riemannian manifolds and we describe the Fubini-Study metric tensor.
    \item In section \ref{QGT_complex_params} we derive the metric tensor with respect to complex parameters and state our main results.
    \item In section \ref{Numerical_experiments} we present and discuss some numerical simulations.
\end{itemize}

\section{Review of Natural Gradient}
\label{Previously}

\subsection{Steepest descent on Riemannian manifolds}
\label{subsection:steepest}
Let us define a differentiable loss function $L:\mathbb{R}^n \rightarrow \mathbb{R}$ which maps a vector of real parameters $\bm{\theta} = [\theta_1, ..., \theta_n]^T \in \mathbb{R}^n$ to a value in $\mathbb{R}$. One may want to minimize $L(\bm\theta)$ with respect to $\bm{\theta}$. For this purpose, a simple optimization algorithm is gradient descent, which is based on a first order expansion of $L$ (throughout our manuscript we use the Einstein summation convention, i.e. repeated indices are summed over):
\begin{align}
\label{L_first_order_dvpt}
    L(\bm\theta + d\bm\theta) \approx L(\bm{\theta}) +
    \frac{\partial L(\bm\theta)}{\partial\theta_i}d\theta_i.
\end{align}
Therefore, a simple minimization algorithm is given by the update rule (\ref{1st_update_rule}), where $\eta_t \in \mathbb{R}$ is called the learning rate and may depend on the step $t$:
\begin{align}
    \label{1st_update_rule}
    \bm{\theta}_{t+1} = \bm{\theta}_t - \eta_t\partial_{\bm\theta}L_t,
\end{align}
where we defined $\partial_{\bm\theta}L_t=\frac{\partial L(\bm\theta_t)}{\partial \bm{\theta}}$.
In Eq.~\eqref{1st_update_rule}, each direction is weighted equally by a uniform learning rate $\epsilon_t$, which is fine if infinitesimal distances are computed in a Euclidean way:
\begin{align}
    d_E(\bm\theta, \bm\theta+d\bm\theta) = \sqrt{d\theta_i d\theta_i}.
\end{align}
However, if distances are computed using a different metric, the gradient computed from the loss function as in Eq.~\eqref{1st_update_rule} does no longer point in the optimal direction anymore. Let us then endow the parameter space with a metric tensor $g(\bm\theta)$ such that an infinitesimal length at the point $\bm\theta$ is computed as:
\begin{align}
    d(\bm\theta,\bm\theta+d\bm\theta) = \sqrt{ g(\bm\theta)_{ij} \mathrm{d}\theta_i \mathrm{d} \theta_j}
\end{align} (Eq.~\eqref{1st_update_rule} corresponds to $g(\bm\theta) = \mathbb{1}$).
We now have a Riemannian structure on the parameter space, characterized by the metric tensor $g(\bm\theta)$ that warps the space. From an infinitesimal perspective, the steepest ascent/descent direction is going to be adjusted by $g$. From a macroscopic perspective, distances  between  points  must  be computed by integrating along a geodesic, with the consequence that the shortest path between two points is generally no longer a straight line as in Euclidean space. Note that from now on we will omit the dependence on $\bm{\theta}$ when writing $g$, but we stress that $g$ is generally a local quantity.

It can then be shown using Lagrange multipliers \cite{amari1998natural} that the steepest descent direction on a Riemannian manifold is given by the so-called Natural Gradient, defined as:
\begin{align}
    \nabla_{\bm{\theta}} L := g^{-1} \partial_{\bm{\theta}} L
\end{align}

This leads to a new parameter update rule:
\begin{align}
    \label{2nd_update_rule}
    \bm{\theta}_{t+1} &= \bm{\theta}_t - \eta_t\nabla_{\bm\theta}L_t\\
    &=\bm{\theta}_t - \eta_tg^{-1}\partial_{\bm\theta}L_t
\end{align}

Note that the metric tensor $g$ may happen to be non-invertible at some values of $\bm{\theta}$. This corresponds to singular points in the parameter space and will be discussed briefly later on.

As an example we can present polar coordinates for the two-dimensional plane  \cite{amari1998why}. When using $(r, \phi) \in \mathbb{R}_+ \times [0, 2\pi[$ to identify points, one must remember that $\phi$ represents an angle, and therefore not a distance per se. Thus, if one moves away from the point $(r, \phi)$ with infinitesimal shifts $\mathrm{d}r$ and $\mathrm{d}\phi$, the infinitesimal distance will not be computed as:
\begin{align}
    \label{false_dist_polar}
    \cancel{\mathrm{d}s = \sqrt{\mathrm{d}r^2 + \mathrm{d}\phi^2}}
\end{align}
but as:
\begin{align}
    \label{inf_dist_polar}
    \mathrm{d}s = \sqrt{\mathrm{d}r^2 + r^2 \mathrm{d}\phi^2}
\end{align}
Thus, the metric tensor for polar coordinates is:
\begin{align}
    \label{metric_polar}
    g = 
    \begin{pmatrix}
         1 & 0 \\
         0 & r^2
    \end{pmatrix}
\end{align}
The metric is singular at the origin, where it becomes non-invertible. This does not mean that infinitesimal distances from the origin cannot be calculated, it simply means that at least one parameter becomes redundant (in fact at the origin the second term in Eq.~\eqref{inf_dist_polar} goes to zero and $ds = dr$), which justifies the use of pseudo-inverses.

The natural gradient for a loss function defined on polar coordinates then becomes 
\begin{align}
\nabla_{(r,\phi)}L = 
    \begin{pmatrix}
         \frac{\partial L}{\partial r}\\
         \frac{1}{r^2}\frac{\partial L}{\partial\phi}\\
    \end{pmatrix}
\end{align}

\subsection{The Fubini-Study metric tensor}

Let us consider a parametrized quantum state $\psi_{\bm\theta}$ where the parameters are real, i.e. $\bm\theta \in\mathbb{R}^n$:
\begin{align}
    \label{ansatz}
    \psi_{\bm\theta} = U(\bm{\theta})\psi_{\bm 0}
\end{align}
where $U(\bm{\theta})$ denotes a parametrized unitary transformation, and $\psi_{\bm 0}$ a fixed initial state. Typically, such a state can be used to describe an ansatz prepared by a parametrized quantum circuit \cite{koczor2020quantum, Straaten2020measurement}.

In a VQE context, a Hamiltonian $\mathcal{H}$ is eventually measured. In this case, the loss function represents the energy, which one aims at minimizing:
\begin{equation}
    \label{energy}
    L(\bm{\theta}) := \langle\psi_{\bm\theta},\mathcal{H}\psi_{\bm\theta}\rangle
\end{equation}
Otherwise, one could use quantum fidelity for the purpose of state preparation or gate synthesis, which one seeks to maximize  \cite{Arrazola_2019, miatto2020fast, yuan2021fast}. Fidelity is a measure of similarity between the current output state, $\psi_{\bm\theta}$, and a target state $\psi_\mathrm{target}$. If the states are both pure, then quantum fidelity is the square of the overlap between the two states and the loss function therefore is:
\begin{equation}
    \label{eq:fidelityloss}
    L(\bm{\theta}) = - |\langle\psi_\mathrm{target},\psi_{\bm\theta}\rangle|^2
\end{equation}
Note that minimization of this loss function is equivalent to a minimization of energy where the Hamiltonian is $\mathcal{H} = - \psi_\mathrm{target}\psi_\mathrm{target}^\dagger$. Rather conveniently, fidelity is not sensitive to the global phase of $\psi_{\bm\theta}$ or $\psi_\mathrm{target}$, which is a useful feature, given that global phase is physically unobservable. Sensitivity to global phase (as we will see shortly) would be an issue of using Euclidean distance as a loss, such as $\lVert \psi_\mathrm{target} - \psi_{\bm\theta} \rVert$.

Following the approach of Provost and Vallee \cite{provost1980Riemannian}, we consider the infinitesimal Euclidean distance between two states and then we adjust it to be insensitive to  global phase.
We start from a first order expansion of $\psi_{\bm\theta}$ around $\bm\theta$:
\begin{align}
    \label{psi_expansion}
        \psi_{\bm\theta + d\bm\theta} & \approx \psi_{\bm\theta} +  \frac{\partial \psi_{\bm\theta}}{\partial \theta_i} d\theta_i\\
        &=\psit + \partial_i\psit d\theta_i
\end{align}
where we introduced the simplified notation $\partial_i = \frac{\partial}{\partial\theta_i}$. Therefore the Euclidean distance between $\psi_{\bm\theta}$ and $\psi_{\bm\theta + d\bm\theta}$ is
\begin{align}
    ds^2=\lVert \psi_{\bm\theta} - \psi_{\bm\theta+d\bm\theta}\rVert^2 &= \left\langle\partial_i\psit,\partial_j\psit\right\rangle d\theta_id\theta_j
\end{align}
Separating the real and imaginary parts of the Hermitian tensor we can write
\begin{align}
    \left\langle\partial_i\psit,\partial_j\psit\right\rangle = \gamma_{ij} + i\sigma_{ij}
\end{align}
where $\gamma_{ij} = \gamma_{ji}$ and $\sigma_{ij} = -\sigma_{ji}$ due to the inner product having to satisfy $\langle a,b\rangle = \langle b,a\rangle^*$. This implies that effectively only the real part matters (as the full contraction of a symmetric tensor with an antisymmetric one yields zero), i.e. 
\begin{align}
    \lVert \psi_{\bm\theta} - \psi_{\bm\theta+d\bm\theta}\rVert^2 &= \gamma_{ij} d\theta_id\theta_j
\end{align}
The antisymmetric part $\sigma_{ij} = \mathrm{Im}[\langle\partial_i\psit,\partial_j\psit\rangle]$ is known as the Berry connection \cite{Wilczek1989tensorandberry}, and its significance is beyond the scope of this work.

At this point, we cannot interpret $\gamma_{ij}$ as a metric tensor on the space of physical quantum states because it is sensitive to global phase: had we used the physically identical state $\psit' = e^{i\alpha(\bm\theta)}\psit$ for some real function $\alpha(\bm\theta)$, the tensor $\gamma_{ij}$ would have been
\begin{align}
    \gamma_{ij}'&=\mathrm{Re}\left[\left\langle\partial_i\psit,\partial_j\psit\right\rangle\right]\\
    \label{eq:newgamma}
    &=\gamma_{ij} +(\partial_i{\alpha})\beta_j + (\partial_j{\alpha})\beta_i+
    (\partial_i{\alpha})(\partial_j{\alpha})
\end{align}
with $\beta_j = -i\langle\psit,\partial_j{\psit}\rangle$. Note that $\beta_j\in\mathbb{R}$ due to the norm of $\psit$ being a constant:
\begin{align}
    &0 = \deriv{}{\theta_j}\left\langle\psit,\psit\right\rangle = \left\langle\psit,\partial_j{\psit}\right\rangle + \left\langle\partial_j{\psit},\psit\right\rangle\\
    &\Rightarrow \left\langle\psit,\partial_j{\psit}\right\rangle = -\left\langle\partial_j{\psit},\psit\right\rangle= -\left\langle\psit,\partial_j{\psit}\right\rangle^* 
\end{align}
Under a change in global phase, we have $\beta_j\rightarrow\beta_j + \partial_j{\alpha}$, which together with Eq.~\eqref{eq:newgamma} implies that the real, symmetric, positive definite tensor \begin{align}
    g_{ij} &= \gamma_{ij} - \beta_i\beta_j\\
    \label{FSmetric}
    &=\mathrm{Re}\left[\left\langle\partial_i{\psit},\partial_j{\psit}\right\rangle\right] - \left\langle\partial_i{\psit},\psit\right\rangle\left\langle\psit,\partial_j{\psit}\right\rangle
\end{align}
is invariant under changes in global phase. The tensor $g_{ij}$ is known as the Fubini-Study (FS) metric.

An alternative way of obtaining the FS metric is to start from a distance measure that is already invariant under a change in global phase, such as 1 minus the fidelity: 
\begin{align}
ds^2&=1-|\langle\psit,\psidt\rangle|^2\\
&=1-|1+\LRangle{\psit,\partial_i\psit} d\theta_i+\frac{1}{2}\LRangle{\psit,\partial_i\partial_j\psit} d\theta_i d\theta_j|^2\\
&=-\bigg(\frac12\LRangle{\psit,\partial_i\partial_j\psit}+\frac12\LRangle{\partial_i\partial_j\psit,\psit}\nonumber\\
&\hspace{5em}+\LRangle{\partial_i\psit,\psit}\LRangle{\psit,\partial_j\psit}\bigg)d\theta_i d\theta_j\\
&=\biggl(\mathrm{Re}\bigl[\LRangle{\partial_i\psit,\partial_j\psit}\bigr] - \LRangle{\partial_i\psit,\psit}\LRangle{\psit,\partial_j\psit}\biggr)d\theta_i d\theta_j
\end{align}
Where we used a second order expansion, and where the last step is due to the identities
\begin{align}
\label{eq:id1}
    \partial_i\langle\psit,\partial_j\psit\rangle &= \langle\partial_i\psit,\partial_j\psit\rangle + \langle\psit,\partial_i\partial_j\psit\rangle\\
\label{eq:id2}
    \partial_i\langle\partial_j\psit,\psit\rangle &= \langle\partial_i\partial_j\psit,\psit\rangle + \langle\partial_j\psit,\partial_i\psit\rangle
\end{align}
and therefore, given that $ \partial_i\langle\psit,\partial_j\psit\rangle = -\partial_i\langle\partial_j\psit,\psit\rangle$, if we sum \eqref{eq:id1} and \eqref{eq:id2} we obtain
\begin{align}
    \langle\psit,\partial_i\partial_j\psit\rangle + \langle\partial_i\partial_j\psit,\psit\rangle &= -\langle\partial_i\psit,\partial_j\psit\rangle - \langle\partial_j\psit,\partial_i\psit\rangle\nonumber\\
    &= -2\mathrm{Re}[\langle\partial_i\psit,\partial_j\psit\rangle]
\end{align}

\section{Complex Natural Gradient for quantum optical circuits}
\label{QGT_complex_params}

We now generalize the FS metric to the case where the parametrization of $\psi$ can involve a mix of real and complex parameters, and we apply it to the optimization of quantum optical circuits.

We will achieve our goal in two steps: first, we will convert complex parameters into their real and imaginary parts; second, we will reassemble the two parts by applying the rules of Wirtinger calculus \cite{Wirtinger1927, hunger2007introduction, kreutz2009complex} to rewrite the Geometric tensor in terms of derivatives with respect to complex parameters and their conjugate as \emph{independent} variables. This approach will make it straightforward to deal with non-holomorphic loss functions.

\subsection{Real and complex parameters}
A complex parameter $z$ and its conjugate $z^*$ are related to the real and imaginary parts of $z$ by a linear transformation $W$: 
\begin{align}
\label{complextoreal}
    \begin{pmatrix}
    z\\z^*
    \end{pmatrix}&=
    \begin{pmatrix}
    1 & i\\
    1 & -i
    \end{pmatrix}
    \begin{pmatrix}
    z_R\\z_I
    \end{pmatrix}=W \begin{pmatrix}
    z_R\\z_I
    \end{pmatrix}\\
    \begin{pmatrix}
    z_R\\z_I
    \end{pmatrix}&=
    \frac{1}{2}\begin{pmatrix}
    1 & 1\\
    -i & i
    \end{pmatrix}
    \begin{pmatrix}
    z\\z^*
    \end{pmatrix}=W^{-1}\begin{pmatrix}
    z\\z^*
    \end{pmatrix},
\end{align}
where $z_R = \mathrm{Re}(z)$ and $z_I = \mathrm{Im}(z)$.
Similarly, the gradients with respect to $z, z^*$ and the gradients with respect to $z_R, z_I$ are related by a linear transformation $V$, which we can find by applying the chain rule:
\begin{align}
\label{eq:chainrule}
    \frac{\partial \psi}{\partial z_R} &= \frac{\partial \psi}{\partial z}\frac{\partial z}{\partial z_R} + \frac{\partial \psi}{\partial z^*}\frac{\partial z^*}{\partial z_R} = \frac{\partial\psi}{\partial z} + \frac{\partial\psi}{\partial z^*} \\
    \frac{\partial \psi}{\partial z_I} &= \frac{\partial \psi}{\partial z}\frac{\partial z}{\partial z_I} + \frac{\partial \psi}{\partial z^*}\frac{\partial z^*}{\partial z_I} = i\left(\frac{\partial\psi}{\partial z} - \frac{\partial\psi}{\partial z^*}\right),
\end{align}
i.e. $\partial_{z_R} = \partial_z + \partial_{z^*}$ and $\partial_{z_I} = -i(\partial_z - \partial_{z^*})$.
Conversely, $\partial_{z} = \frac12(\partial_{z_R} - i\partial_{z_I})$ and $\partial_{z^*} = \frac12(\partial_{z_R} + i\partial_{z_I})$, which means
\begin{align}
    \begin{pmatrix}
    \partial_z\\\partial_{z^*}
    \end{pmatrix}&=
    \frac12\begin{pmatrix}
    1 & -i\\
    1 & i
    \end{pmatrix}
    \begin{pmatrix}
    \partial_{z_R}\\\partial_{z_I}
    \end{pmatrix}=V\begin{pmatrix}
    \partial_{z_R}\\\partial_{z_I}
    \end{pmatrix}\\
    \begin{pmatrix}
    \partial_{z_R}\\\partial_{z_I}
    \end{pmatrix}&=
    \begin{pmatrix}
    1 & 1\\
    i & -i
    \end{pmatrix}
    \begin{pmatrix}
    \partial_z\\\partial_{z^*}
    \end{pmatrix}=V^{-1}\begin{pmatrix}
    \partial_z\\\partial_{z^*}
    \end{pmatrix}.
\end{align}
In the subsection that follows, we will write $V$ and $W$ even for larger collections of parameters, with the understanding that $V$ and $W$ will be block-diagonal with $2\times2$ blocks for complex parameters and if a parameter is real its block will be $1\times1$ with value 1.

\subsection{The FS metric with respect to parameters of any type}
In this section, we will derive an expression of the \emph{hermitian} FS metric $f$ for parameters of any type.

Instead of writing the FS metric tensor using the real part function (which is not holomorphic), we write it using the fact that the FS metric is the symmetric part of the Geometric tensor $G_{ij}$:
\begin{align}
    g &= \frac{G + {G^T}}{2},
\end{align}
where
\begin{align}
\label{GRformula}
G_{ij} &= \left\langle\frac{\partial \bm{\psi}}{\partial \theta_i},\frac{\partial \bm{\psi}}{\partial \theta_j}\right\rangle - \left\langle\frac{\partial \bm{\psi}}{\partial \theta_i}, \bm{\psi}\right\rangle\left\langle\bm{\psi},\frac{\partial \bm{\psi}}{\partial \theta_j}\right\rangle.
\end{align}

Using such linear relation between $g$ and $G$, we can write the FS metric with respect to a mixture of real and complex parameters. Define $\bm\xi$ as our parameter vector, containing real parameters and complex parameters with the provision that for each complex parameter we also include its complex conjugate. In this way, the all-real Natural Gradient update rule
\begin{align}
\label{greal}
    \bm{\theta}\leftarrow \bm{\theta} - \epsilon g^{-1}\frac{\partial L}{\partial \bm{\theta}}
\end{align}
turns into the more general update rule
\begin{align}
\label{fcomplex}
    \bm{\xi}\leftarrow \bm{\xi} - \epsilon f^{-1}\frac{\partial L}{\partial \bm{\xi}^*},
\end{align}
which works for any type of parameter, real or complex.

Note that the gradient in the update rule for a complex parameter $\xi_i$ is written with respect to the \emph{conjugate} of the parameter \cite{hunger2007introduction}. Such update obviously falls back to the usual one in case $\xi_i$ is real.

We derive the FS metric $f$ in two steps. First, we write 
\begin{align}
    \frac{\partial L}{\partial \bm{\theta}} = {V^{-1}}^*\frac{\partial L}{\partial\bm{\xi}^*} = W^\dagger\frac{\partial L}{\partial\bm{\xi}^*},
\end{align}
where we used the functional relation $V^{-1}=W^T$. Second, we transform the basis from $\bm{\theta}$ to $\bm{\xi}$ in Eq.~\eqref{greal} using Eq.~\eqref{complextoreal} to obtain the gradient updates for $\bm{\xi}$:
\begin{align}
    \label{transformed}
    \bm{\xi}\leftarrow \bm{\xi} - \epsilon Wg^{-1}{W^{\dagger}}\frac{\partial L}{\partial \bm{\xi}^*}.
\end{align}
So we deduce by comparing \eqref{fcomplex} and \eqref{transformed} that the metric tensor with respect to the complex parameters is given by
\begin{align}
    f &= (Wg^{-1}{W^{\dagger}})^{-1} = (W^{\dagger})^{-1}gW^{-1}\\
    &=\frac{(W^{\dagger})^{-1}GW^{-1} + (W^{\dagger})^{-1}{G}^TW^{-1}}{2}\\
    &=\frac{V^*GV^T + V^*{G}^TV^T}{2}.
\label{fformula0}
\end{align}
In the last step we used the functional relation $W^{-1}=V^T$. Finally, we find the expression for $f_{mn}$ by inserting Eq.~\ref{GRformula} into Eq.~\ref{fformula0}:
\begin{align}
    \hspace{-10em}f_{mn} 
    &=\frac{1}{2}\left\langle V_{mi}\frac{\partial \bm{\psi}}{\partial \theta_i},V_{nj}\frac{\partial \bm{\psi}}{\partial \theta_j}\right\rangle - \frac{1}{2}\left\langle V_{mi}\frac{\partial \bm{\psi}}{\partial \theta_i}, \bm{\psi}\right\rangle\left\langle\bm{\psi},V_{nj}\frac{\partial \bm{\psi}}{\partial \theta_j}\right\rangle \nonumber\\
    &+\frac{1}{2}\left\langle V^*_{nj}\frac{\partial \bm{\psi}}{\partial \theta_j},W^*_{mi}\frac{\partial \bm{\psi}}{\partial \theta_i}\right\rangle  - \frac{1}{2}\left\langle V^*_{nj}\frac{\partial \bm{\psi}}{\partial \theta_j}, \bm{\psi}\right\rangle\left\langle\bm{\psi},W^*_{mi}\frac{\partial \bm{\psi}}{\partial \theta_i}\right\rangle \\
    \label{fexpression}
    &=\frac{1}{2}\left\langle \frac{\partial \bm{\psi}}{\partial \xi_m},\frac{\partial \bm{\psi}}{\partial \xi_n}\right\rangle - \frac{1}{2}\left\langle \frac{\partial \bm{\psi}}{\partial \xi_m}, \bm{\psi}\right\rangle\left\langle\bm{\psi},\frac{\partial \bm{\psi}}{\partial \xi_n}\right\rangle  \nonumber\\
    &+\frac{1}{2}\left\langle \frac{\partial \bm{\psi}}{\partial \xi_n^*},\frac{\partial \bm{\psi}}{\partial \xi_m^*}\right\rangle - \frac{1}{2}\left\langle \frac{\partial \bm{\psi}}{\partial \xi_n^*}, \bm{\psi}\right\rangle\left\langle\bm{\psi},\frac{\partial \bm{\psi}}{\partial \xi_m^*}\right\rangle.
\end{align}
Note that for real parameters, the expression for $f$ in Eq.~\eqref{fexpression} falls back to the usual FS metric in \eqref{FSmetric}.
Also note that the tensor $f$ is hermitian rather than symmetric (i.e.~$f^T = f^*$), which is to be expected, as the metric tensor of a complex manifold is hermitian.
\section{Application to quantum state preparation}
\label{Numerical_experiments}

We implemented the Natural Gradient descent algorithm in our differentiable simulator of quantum optical circuits \emph{poenta} \cite{miatto2020fast,yuan2021fast}, and used it to train a single-mode quantum circuit for generating a single-photon state and a Hex-GKP state.

Our circuit architecture is made of $N$ layers, each implementing a generic Gaussian transformation $G$ followed by a non-Gaussian transformation $K$:
\begin{align}
    U = K_N G_N K_{N-1}G_{N-1}\dots K_1 G_1
\end{align}
In particular, a single-mode generic Gaussian transformation can be parametrized as $G(\gamma,\phi,\zeta) = D(\gamma)R(\phi)S(\zeta)$, i.e. as a squeezing gate $S(\zeta)=e^{\frac{\zeta^*}{2}a^2-\frac{\zeta}{2}{a^\dagger}^2}$ followed by a phase rotation gate $R(\phi)=e^{i\phi a^\dagger a}$ and a displacement gate $D(\gamma)=e^{\gamma a^\dagger - \gamma^* a}$, where $a^\dagger$ and $a$ denote the annihilation operator and creation operator. Such parametrization can be generalized to generic multi-mode Gaussian gates by employing beam-splitter gates as elements that couple pairs of modes (more details in Ref.~\cite{miatto2020fast}). The non-Gaussian transformation is a Kerr interaction $K(\kappa) = e^{i\kappa (a^\dagger a)^2}$. Note that $\zeta$ and $\gamma$ are complex parameters, while $\phi$ and $\kappa$ are real. Our circuit therefore is parametrized by a mix of real and complex parameters.

We set up a state preparation task where we initialize the circuit parameters randomly and we evaluate the output of the circuit via the fidelity to a target state. The task is then expressed as a minimization of a loss function:
\begin{align}
    \min_{\bm\xi} L(\bm\xi)
\end{align}
where $L(\bm\xi)=1-|\langle\psi_\mathrm{target}|U_{\bm\xi}|0\rangle|^2$ and where $\bm\xi$ represents the entire collection of real and complex parameters of the circuit.

There are two ways to compute the Natural Gradient: the first is with respect to all the parameters in the entire circuit at once, the second is with respect to the parameters of the layer being updated. In the first case we compute the gradient of the loss function with respect to the entire collection of parameters of the circuit, then we compute the full Hermitian metric tensor, we invert it and finally we apply it to the full gradient vector before performing a single gradient descent step. This can be very costly if the number of layers is large, as the effective number of parameters (i.e. counting double for complex ones) in each $N$-mode layer is $n = 2N^2 + 4N$, and the metric tensor is $n\times n$. This gives an estimate on the cost of matrix inversion as $O(n^3) = O(N^6)$. In the second case, we compute the Hermitian metric in block-diagonal form, where we only include in each block the parameters that belong to the same layer. This technique has been studied in \cite{Bromley_2020}, where it was shown that using the block-diagonal metric is in practice just as good as using the full metric while sparing a significant amount of computation. Finally, we note that before the (pseudo)inversion step we add a constant regularization term, as we observed that the metric is often singular \cite{yamamoto2019}:
\begin{align}
    \tilde f^+ = (f + \lambda\mathbb{1})^+
\end{align}
where we set an empirical value of $\lambda\approx0.1$ after assessing the results of numerical experiments.
As mentioned in Sec.~\ref{subsection:steepest}, the singularities in the metric correspond to points in parameter space where at least one of the parameters cannot influence the quantum state (as an intuitive analogy, compare a qubit in the state $|0\rangle$ (on the north pole) with a qubit in the state $|+\rangle = (|0\rangle + |1\rangle)/\sqrt{2}$ (on the equator): the $|+\rangle$ state is going to be very sensitive to a rotation around the $z$ axis, but the $|0\rangle$ state is completely insensitive to it).

In the following experiments we compare SGD, Adam and NGD using the optimal learning rate for each (see appendix for more details on choosing the optimal learning rate). 

\subsection{Single-photon source}

\begin{figure}[ht!]
    \includegraphics[scale = 0.65]{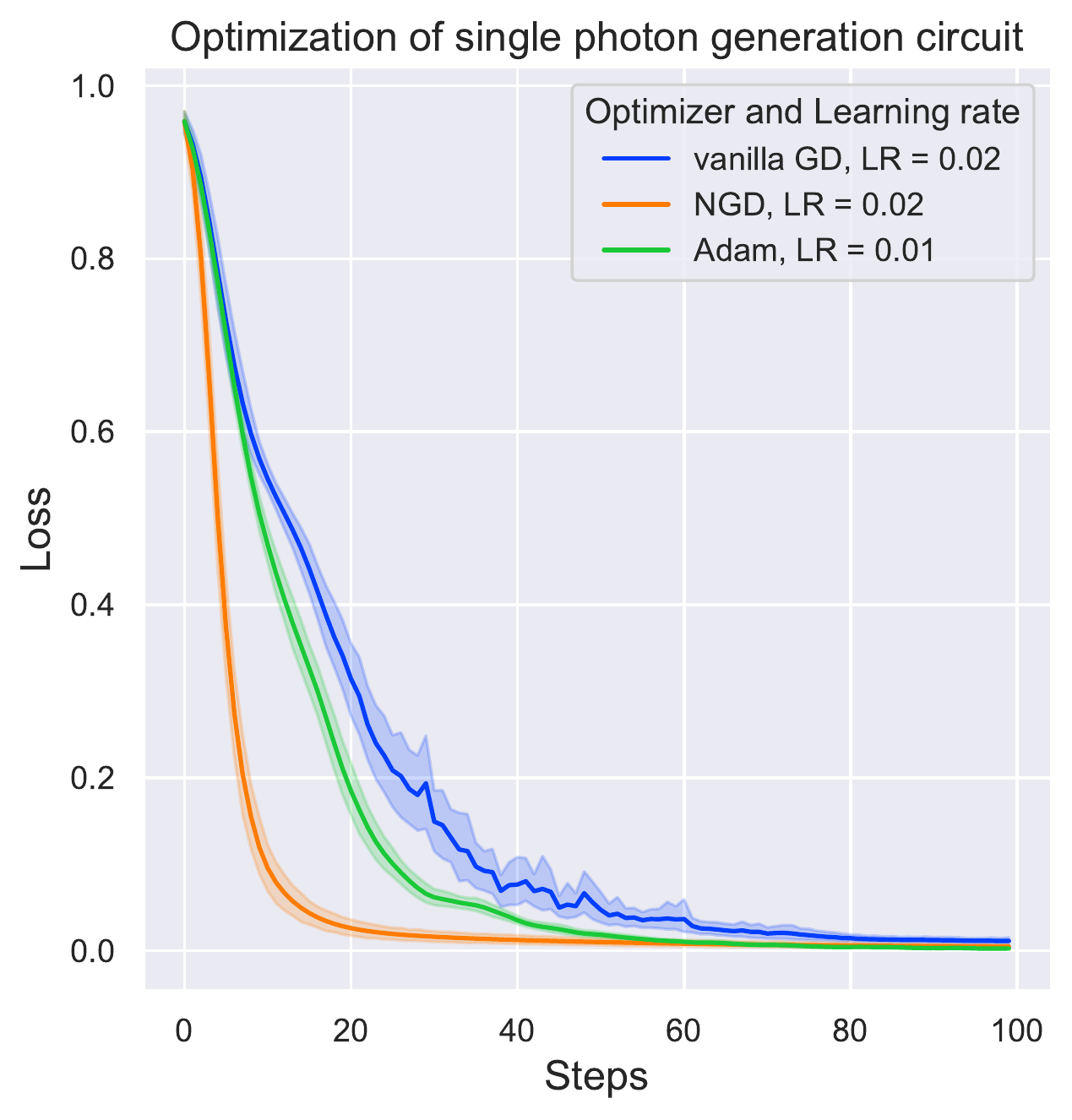}
    \caption{\label{fig:singlephotonNG}Learning a single-photon preparation circuit (8 layers, cutoff dimension of 100). The Natural Gradient converges faster than Adam and SGD, all the while following a smooth decay of the cost function. As this is a single-mode block-diagonal implementation, the dimension of the blocks is at most $6\times 6$, which is so fast to compute that the NG steps have the same runtime as those of Adam and SGD.}
\end{figure}

The first target that we choose is a single-photon state $|\psi_\mathrm{target}\rangle = |1\rangle$. We employ an 8-layer circuit with a total of $8\times6 = 48$ parameters and a Fock-space cutoff of 100. We report the results of the optimization in Fig.~\ref{fig:singlephotonNG}, where one should notice that even for a simple task such as generating a single-photon, the NGD converges significantly faster than Adam.

\subsection{Hex-GKP source}
\begin{figure}[ht!]
    \centering
    \includegraphics[scale=0.65]{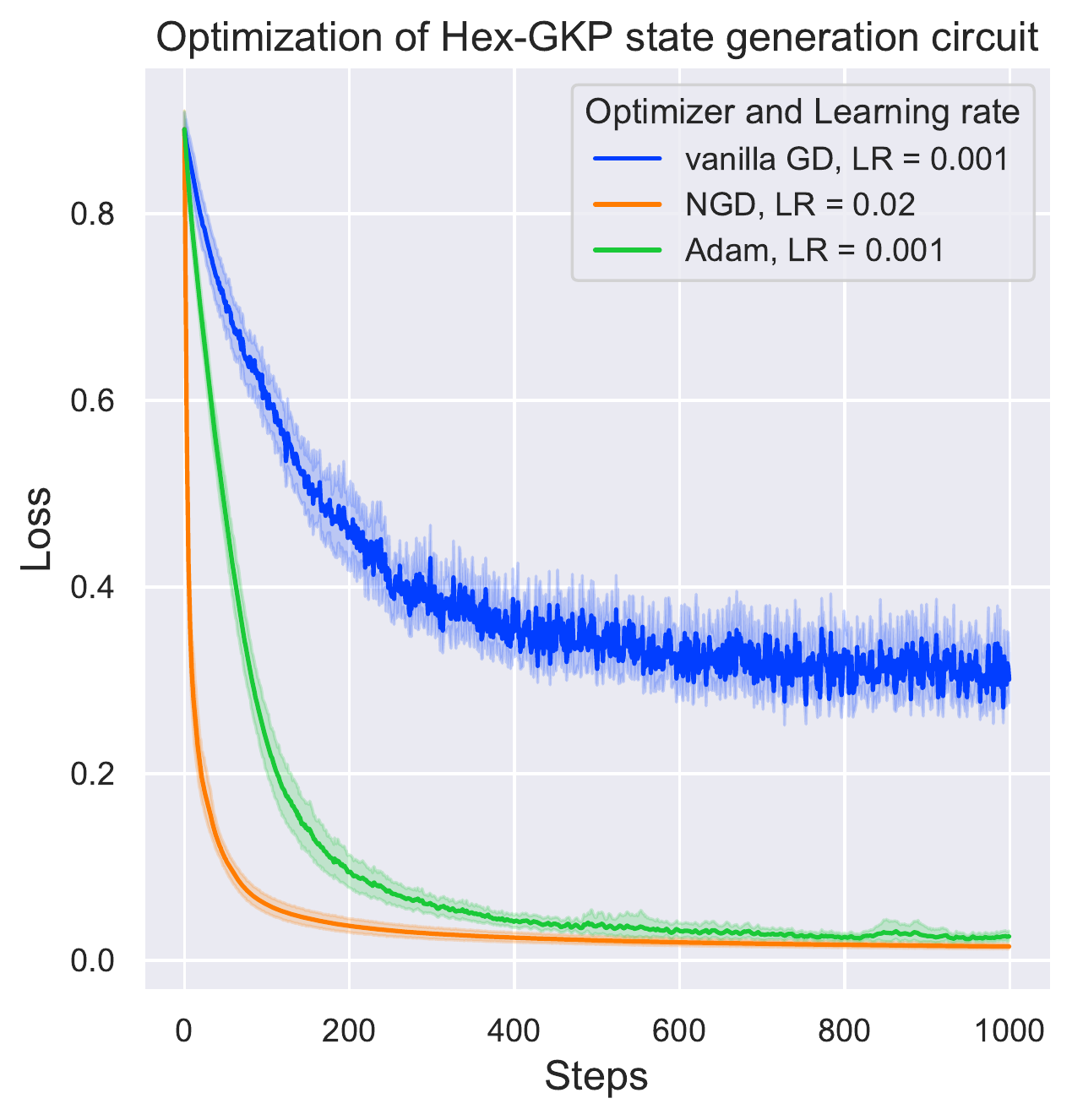}
    \caption{Learning a Hex-GKP preparation circuit (25 layers, cutoff dimension of 50). This is a clear example of the superiority of NGD when it comes to cost functions for difficult state preparation tasks. Not only the runtime is the same as the other two methods (this is still a single-mode circuit), but the NGD follows a much smoother curve and it reaches an overall lower cost.}
    \label{fig:hexGkpNG}
\end{figure}

The second target that we picked is the Hex-GKP state as defined in \cite{Arrazola_2019} with $d=2$, $\mu=1$, $\delta=0.3$. We use a 25-layer circuit with a total of 150 parameters and a Fock-space cutoff of 50. The results are shown in Fig.~\ref{fig:hexGkpNG}. 

This is a much harder task than the previous one and the cost landscape is much more rugged, as it can be seen from the behaviour of the vanilla GD curve. Adam also seems to suffer from the rugged landscape albeit by a much smaller extent. In contrast, the NGD curve is smoother even at a learning rate of 0.02, which allows it to converge very quickly to a high fidelity output.

\section{Conclusions}

In this work we have derived a complex version of the Fubini-Study metric. The resulting Hermitian metric tensor allows us to perform Natural Gradient optimization in a manifold of complex parameters.

We applied this technique to the optimization of quantum optical circuits, showing that it outperforms vanilla gradient descent and Adam, by reaching the minimum of the cost function in fewer steps and in a smoother fashion.

In particular, using NGD we optimized a circuit that produces single photons and a circuit that produces Hex-GKP states in fewer update steps than with other optimization methods, thanks to the fact that inverting the local metric allows us to take steps in a better direction, which in turn allows us to increase the learning rate and therefore take larger steps in the parameter space. Despite taking such large steps, the decay of the cost function is smooth, which indicates that the parameter space itself seen through the inverted metric is itself smoother.

In a future work we will look at pairing the NG with other gradient descent techniques, in particular geodesic optimizations in Riemannian and Hermitian manifolds.

\bibliographystyle{unsrt}
\bibliography{main}

\begin{thebibliography}{10}

\bibitem{Killoran_2019_continuous}
Nathan Killoran, Thomas~R. Bromley, Juan~Miguel Arrazola, Maria Schuld,
  Nicol\'as Quesada, and Seth Lloyd.
\newblock Continuous-variable quantum neural networks.
\newblock {\em Phys. Rev. Research}, 1:033063, Oct 2019.

\bibitem{Arrazola_2019}
Juan~Miguel Arrazola, Thomas~R Bromley, Josh Izaac, Casey~R Myers, Kamil
  Br{\'{a}}dler, and Nathan Killoran.
\newblock Machine learning method for state preparation and gate synthesis on
  photonic quantum computers.
\newblock {\em Quantum Science and Technology}, 4(2):024004, jan 2019.

\bibitem{boyd2020nonlinear}
Robert~W Boyd.
\newblock {\em Nonlinear optics}.
\newblock Academic press, 2020.

\bibitem{bourassa2021fast}
J~Eli Bourassa, Nicol{\'a}s Quesada, Ilan Tzitrin, Antal Sz{\'a}va, Theodor
  Isacsson, Josh Izaac, Krishna~Kumar Sabapathy, Guillaume Dauphinais, and Ish
  Dhand.
\newblock Fast simulation of bosonic qubits via gaussian functions in phase
  space.
\newblock {\em arXiv preprint arXiv:2103.05530}, 2021.

\bibitem{amari1996neural}
Shun-Ichi Amari.
\newblock Neural learning in structured parameter spaces-natural riemannian
  gradient.
\newblock {\em Advances in neural information processing systems}, 9:127--133,
  1996.

\bibitem{amari1998natural}
Shun-Ichi Amari.
\newblock Natural gradient works efficiently in learning.
\newblock {\em Neural computation}, 10(2):251--276, 1998.

\bibitem{amari1998why}
S.~{Amari} and S.~C. {Douglas}.
\newblock Why natural gradient?
\newblock In {\em Proceedings of the 1998 IEEE International Conference on
  Acoustics, Speech and Signal Processing, ICASSP '98 (Cat. No.98CH36181)},
  volume~2, pages 1213--1216 vol.2, 1998.

\bibitem{yamamoto2019}
Naoki Yamamoto.
\newblock On the natural gradient for variational quantum eigensolver.
\newblock {\em arXiv preprint arXiv:1909.05074}, 2019.

\bibitem{haug2021optimal}
Tobias Haug and MS~Kim.
\newblock Optimal training of variational quantum algorithms without barren
  plateaus.
\newblock {\em arXiv preprint arXiv:2104.14543}, 2021.

\bibitem{Stokes2020}
James Stokes, Josh Izaac, Nathan Killoran, and Giuseppe Carleo.
\newblock Quantum {N}atural {G}radient.
\newblock {\em {Quantum}}, 4:269, May 2020.

\bibitem{koczor2020quantum}
B\'{a}lint Koczor and Simon~C. Benjamin.
\newblock Quantum natural gradient generalised to non-unitary circuits.
\newblock {\em arXiv preprint arXiv:1912.08660v4}, 2020.

\bibitem{wierichs2020avoiding}
David Wierichs, Christian Gogolin, and Michael Kastoryano.
\newblock Avoiding local minima in variational quantum eigensolvers with the
  natural gradient optimizer.
\newblock {\em arXiv preprint arXiv:2004.14666}, 2020.

\bibitem{Schuld_2019}
Maria Schuld, Ville Bergholm, Christian Gogolin, Josh Izaac, and Nathan
  Killoran.
\newblock Evaluating analytic gradients on quantum hardware.
\newblock {\em Physical Review A}, 99(3), Mar 2019.

\bibitem{Yuan_2019}
Xiao Yuan, Suguru Endo, Qi~Zhao, Ying Li, and Simon~C. Benjamin.
\newblock Theory of variational quantum simulation.
\newblock {\em Quantum}, 3:191, Oct 2019.

\bibitem{Li_2017_efficient}
Ying Li and Simon~C. Benjamin.
\newblock Efficient variational quantum simulator incorporating active error
  minimization.
\newblock {\em Phys. Rev. X}, 7:021050, Jun 2017.

\bibitem{Preskill_2018}
John Preskill.
\newblock Quantum computing in the nisq era and beyond.
\newblock {\em Quantum}, 2:79, Aug 2018.

\bibitem{Straaten2020measurement}
Barnaby van Straaten and B\'{a}lint Koczor.
\newblock {Measurement cost of metric aware variational quantum algorithms}.
\newblock {\em ArXiv preprint, arXiv:2005.05172v2}, 2020.

\bibitem{Bromley_2020}
Thomas~R Bromley, Juan~Miguel Arrazola, Soran Jahangiri, Josh Izaac,
  Nicol{\'{a}}s Quesada, Alain~Delgado Gran, Maria Schuld, Jeremy Swinarton,
  Zeid Zabaneh, and Nathan Killoran.
\newblock Applications of near-term photonic quantum computers: software and
  algorithms.
\newblock {\em Quantum Science and Technology}, 5(3):034010, may 2020.

\bibitem{miatto2020fast}
Filippo~M. Miatto and Nicolás Quesada.
\newblock Fast optimization of parametrized quantum optical circuits.
\newblock {\em Quantum}, 4:366, Nov 2020.

\bibitem{yuan2021fast}
Yuan Yao and Filippo~M. Miatto.
\newblock Fast differentiable evolution of quantum states under gaussian
  transformations.
\newblock {\em arXiv:2102.05742}, 2021.

\bibitem{provost1980Riemannian}
JP~Provost and G~Vallee.
\newblock Riemannian structure on manifolds of quantum states.
\newblock {\em Communications in Mathematical Physics}, 76(3):289--301, 1980.

\bibitem{Wilczek1989tensorandberry}
F~Wilczek and A~Shapere.
\newblock {\em Geometric Phases in Physics}.
\newblock {WORLD} {SCIENTIFIC}, July 1989.

\bibitem{Wirtinger1927}
W.~Wirtinger.
\newblock Zur formalen theorie der funktionen von mehr komplexen
  veranderlichen.
\newblock {\em Mathematische Annalen}, 97(1):357--375, December 1927.

\bibitem{hunger2007introduction}
Raphael Hunger.
\newblock An introduction to complex differentials and complex
  differentiability.
\newblock 2007.

\bibitem{kreutz2009complex}
Ken Kreutz-Delgado.
\newblock The complex gradient operator and the cr-calculus.
\newblock {\em arXiv preprint arXiv:0906.4835}, 2009.

\end{thebibliography}

\appendix
\section{Optimal learning rate}
\label{appendix:bestlr}
In order to compare the gradient descent algorithms fairly, we cannot use the same learning rate, as it means different things for different algorithms. Moreover, in Adam the learning rate evolves over time.

To determine the best learning rate for each algorithm, we search for it by trial and error as the one that allows the optimizer to reach the lowest value of the cost function quickly and without incurring in excessive oscillations or without blowing up if we let it run.

We run tests separately with the three different algorithms (vanilla GD / SGD, NGD, and Adam) to train the single-mode quantum circuit for generating a single-photon state and a Hex-GKP state. 

For each algorithm, we use 20 random seeds to generate the initial parameters in the circuit and test for a range of learning rates from 0.0001 to 0.5. We present in Fig.~\ref{fig:threealgocomp1} and \ref{fig:threealgocomp2} a few typical curves around the \textit{optimal} learning rate to show how we choose it for each of them.

\onecolumngrid
\begin{center}
\begin{figure}[ht]
    \subfloat[single-photon circuit: Optimal learning rate for vanilla GD is 0.02]{\includegraphics[width=0.33\textwidth]{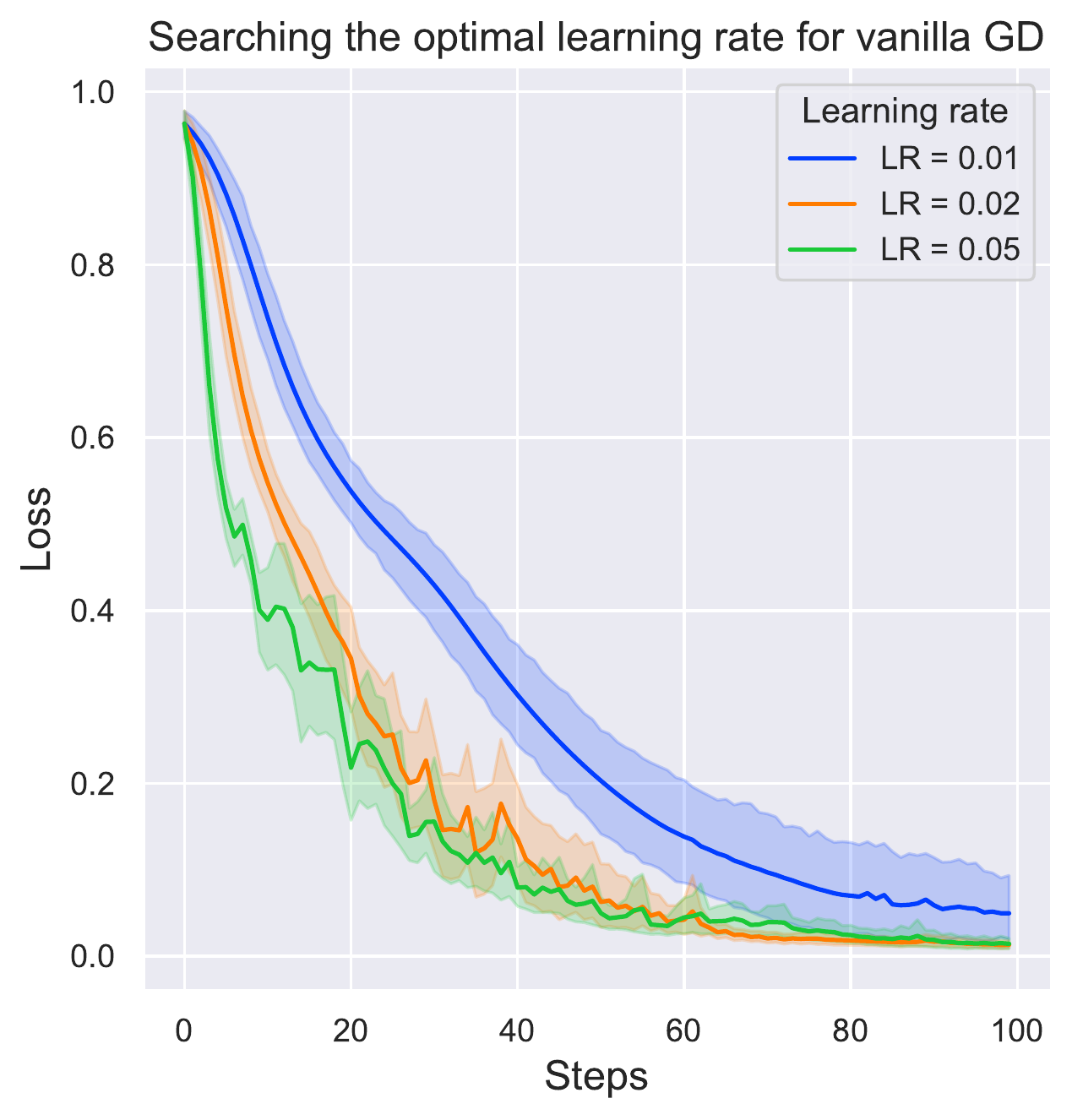}}
    \subfloat[single-photon circuit: Optimal learning rate for NGD is 0.02]{\includegraphics[width=0.33\textwidth]{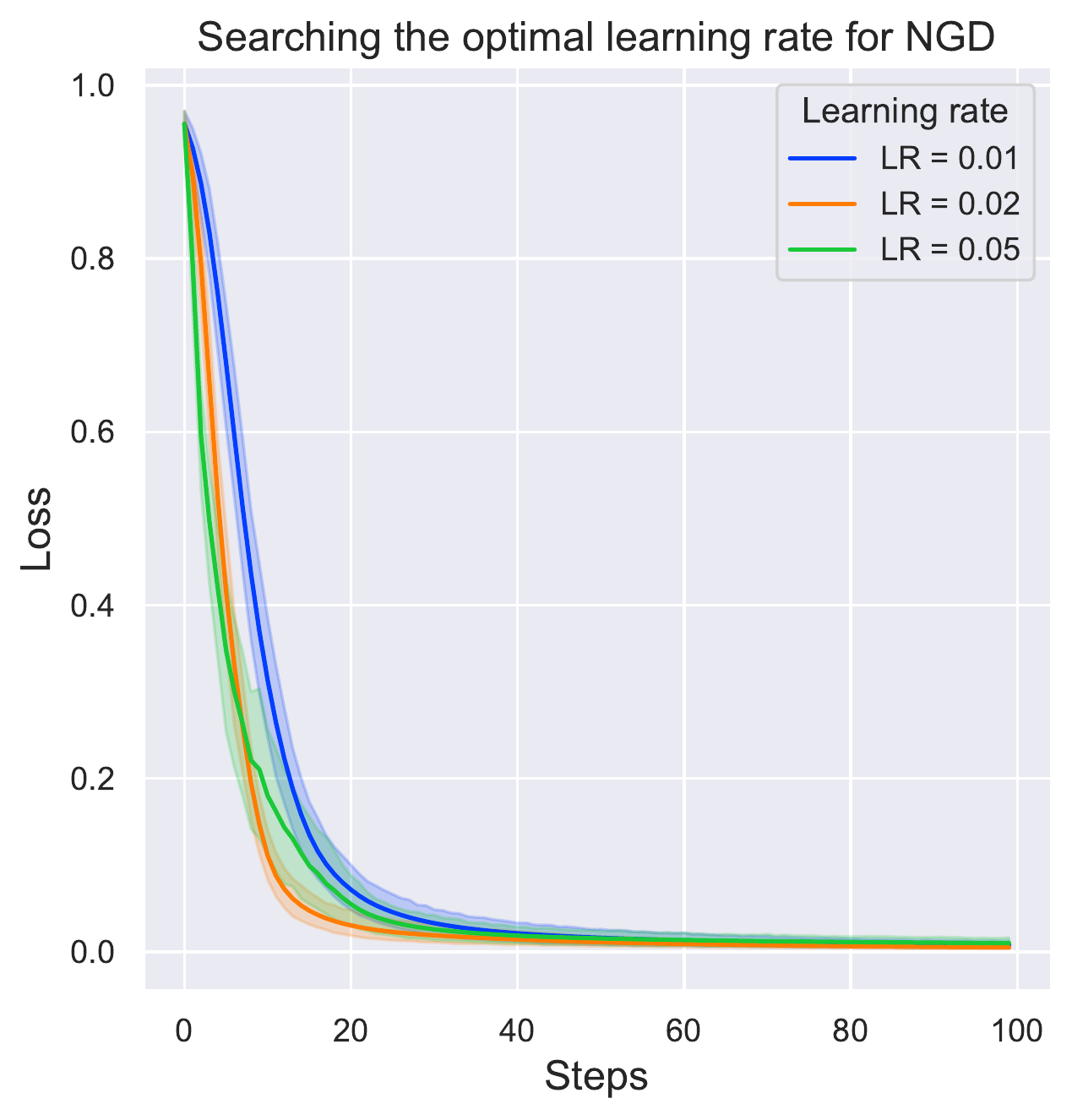}}
    \subfloat[single-photon circuit: Optimal learning rate for Adam is 0.01]{\includegraphics[width=0.33\textwidth]{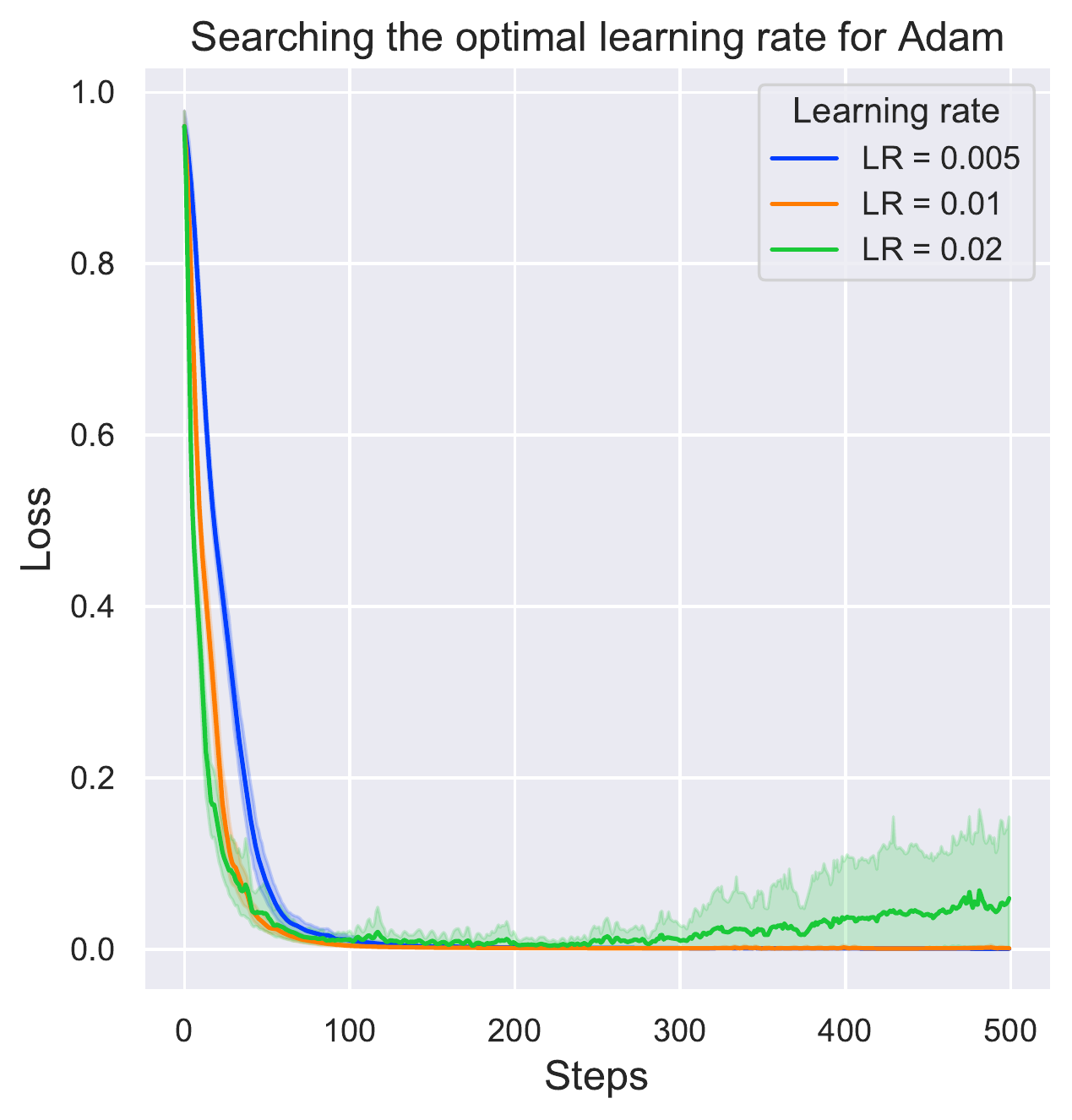}}
    \caption{\label{fig:threealgocomp1}Single-photon preparation circuit. We look for the learning rate that allows the optimizer to reach the lowest value of the cost function quickly and without incurring in excessive oscillations or without blowing up if we let it run.}
\end{figure}
\begin{figure}[ht]
    \subfloat[Hex-GKP circuit: Optimal learning rate for vanilla GD is 0.001]{\includegraphics[width=0.33\textwidth]{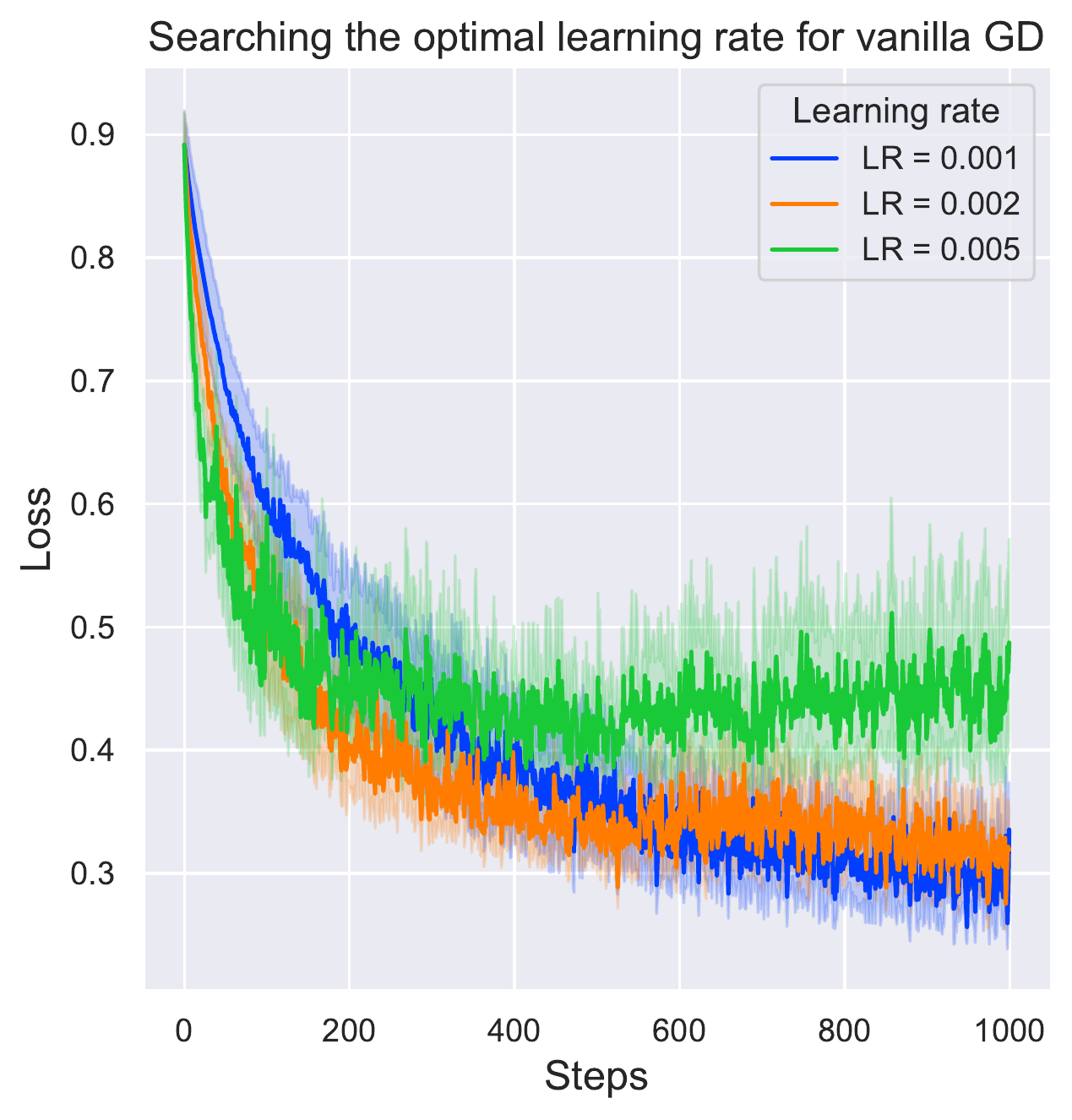}}
    \subfloat[Hex-GKP circuit: Optimal learning rate for NGD is 0.02]{\includegraphics[width=0.33\textwidth]{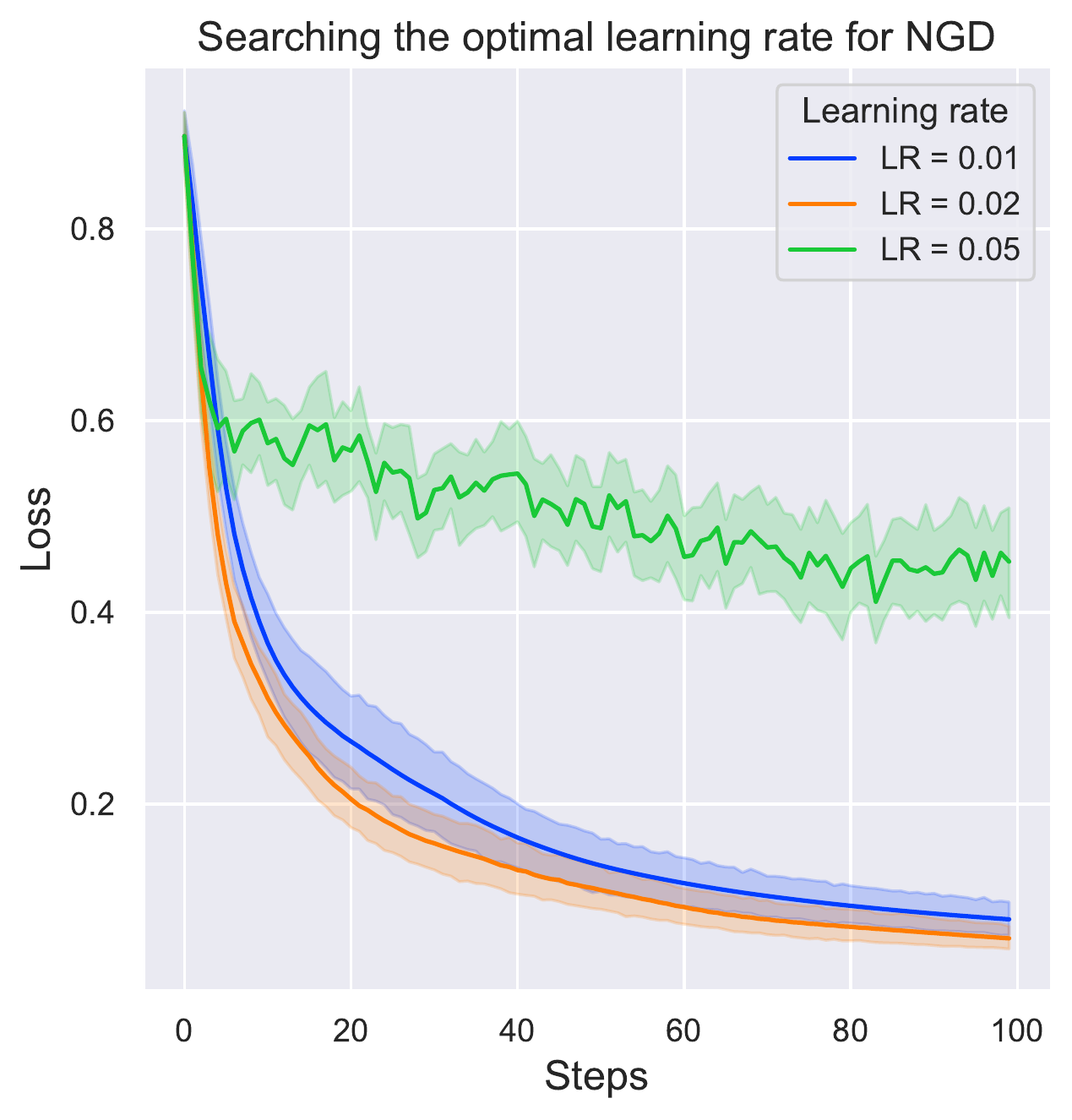}}
    \subfloat[Hex-GKP circuit: Optimal learning rate for Adam is 0.001]{\includegraphics[width=0.33\textwidth]{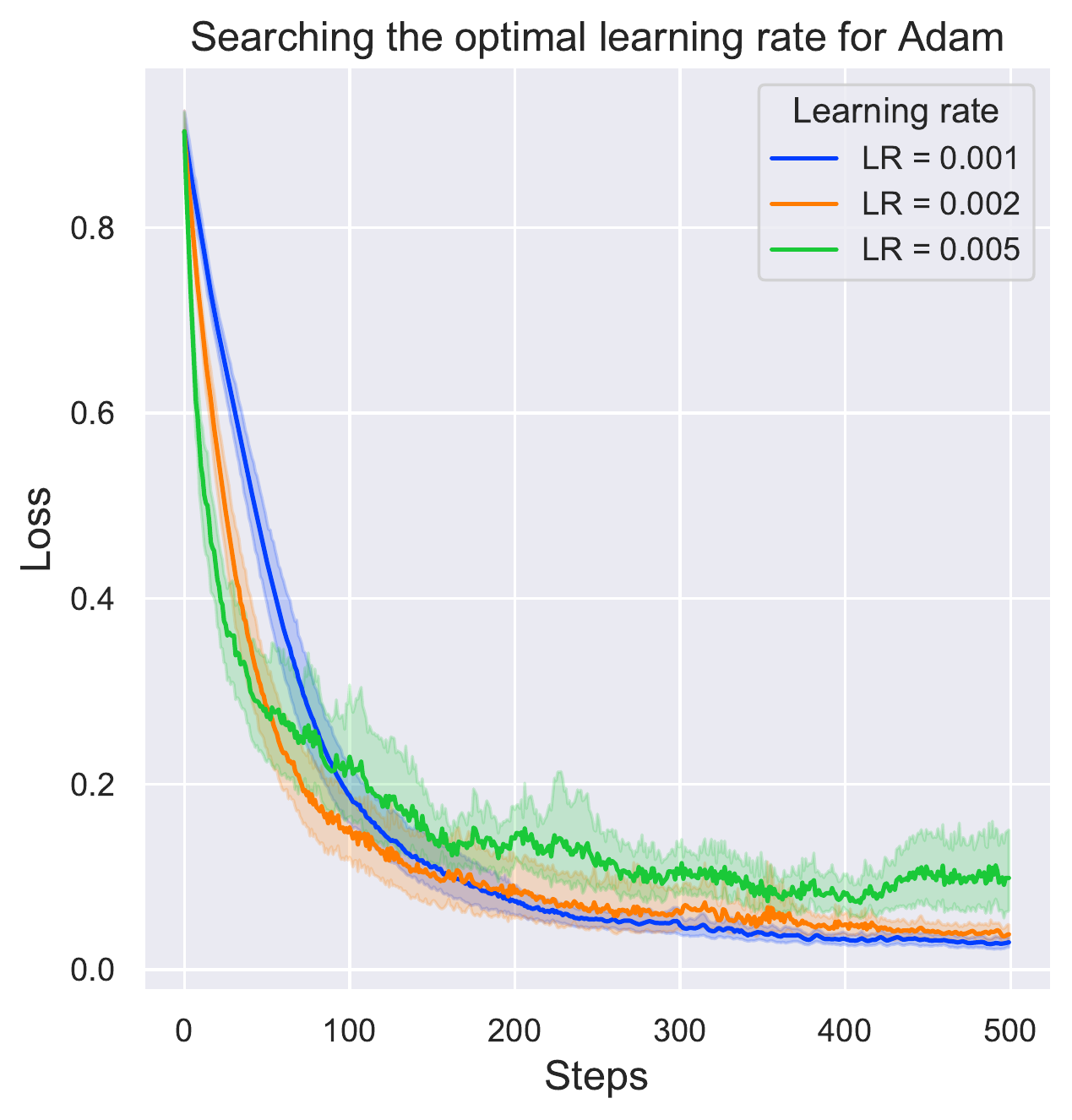}}
    \caption{\label{fig:threealgocomp2}Hex-GKP preparation circuit. We look for the learning rate that allows the optimizer to reach the lowest value of the cost function quickly and without incurring in excessive oscillations or without blowing up if we let it run.}
\end{figure}
\end{center}

\end{document}